\documentclass[prl,aps,twocolumn,superscriptaddress,showpacs]{revtex4}

\usepackage{pxfonts}
\usepackage{graphicx}
\begin{document}

\title{Localization and interaction of indirect excitons in GaAs coupled
quantum wells}

\author{A.\,A. High}
\author{A.\,T. Hammack}
\author{L.\,V. Butov}
\affiliation{Department of Physics, University of California at San
Diego, La Jolla, CA 92093-0319}

\author{L. Mouchliadis}
\author{A.\,L. Ivanov}
\affiliation{Department of Physics and Astronomy, Cardiff
University, Cardiff CF24 3AA, United Kingdom}

\author{M. Hanson}
\author{A.\,C. Gossard}
\affiliation{Materials Department, University of California at Santa
Barbara, Santa Barbara, California 93106-5050}

\begin{abstract}
We introduced an elevated trap technique and exploited it for
lowering the effective temperature of indirect excitons. We observed
narrow photoluminescence lines which correspond to the emission of
individual states of indirect excitons in a disorder potential. We
studied the effect of exciton-exciton interaction on the localized
and delocalized exciton states and found that the homogeneous line
broadening increases with density and dominates the linewidth at
high densities.
\end{abstract}


\date{\today}

\maketitle

Disorder is an intrinsic feature of solid state materials. It forms
due to the spatial inhomogeneity of the sample. In other systems,
such as liquid Helium or cold atomic gases, disorder can be
artificially introduced. The behavior of particles in a disorder
potential is a subject of intensive research. The studies have
concerned electrons in semiconductors and superconductors, liquid
Helium in a porous media, and cold atomic gases in a disorder
potential formed by optical patterning, see e.g.
\cite{Lee85,Ma86,Fisher89,Reppy84,Fallani06}.

Understanding the role of disorder is also essential in the physics
of excitons. For excitons in QW structures, the in-plane disorder
potential forms mainly due to QW width and alloy fluctuations
\cite{Hegarty84,Takagahara85,Zrenner94,Hess94,Gammon96,Zimmermann97}.
Indirect excitons in coupled quantum wells (CQW) composed of an
electron and a hole in separated QWs, Fig.\,1a, form a peculiar
system for studying the effects of disorder. While regular excitons
are neutral particles and interact only weakly, indirect excitons
are electric dipoles with dipole moment $d \approx L_z$, the
distance between the QW centers, and interact as dipoles,
$U_{int}(r)= e^{2}d^{2}/(\epsilon r^{3})$, $r \gg d$
\cite{Yoshioka90,Zhu95,Lozovik96,Ivanov02}. The interaction should
have a significant effect on the localization and transport of
excitons in a disorder. Furthermore, the lifetime of indirect
excitons exceeds by orders of magnitude the lifetime of regular
excitons and indirect excitons can travel over large distances
within their lifetime \cite{Hagn95,Ivanov06,Gartner06} facilitating
the study of exciton transport by imaging spectroscopy. Also, due to
their long lifetime, the indirect excitons can cool to temperatures
close to the lattice temperature \cite{Butov01}, permitting the
study of cold excitons.

In this paper, we introduce an elevated in-plane trap technique
facilitating the observation of individual exciton states in the
disorder potential and present studies of the effect of interaction
on indirect excitons in a disorder.

\begin{figure}
\begin{center}
\includegraphics[width=7cm]{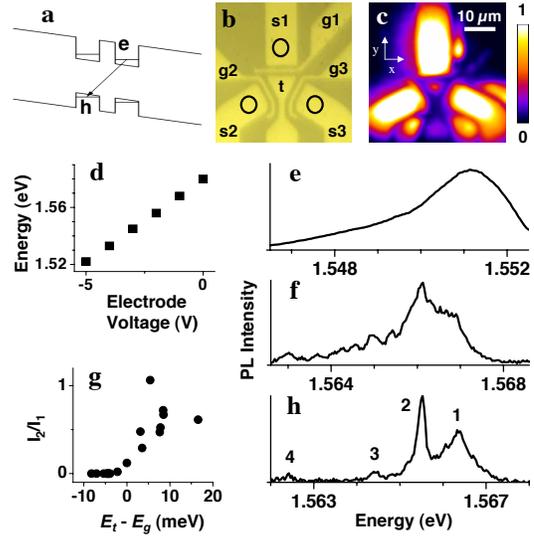}
\caption{(a) Energy band diagram of the CQW structure; e, electron;
h, hole. (b) Electrode pattern. A trap is formed at electrode $t$ by
voltages on electrodes $s$, $g$, and $t$. The circles show the
positions of the laser excitation. (c) Image of the indirect exciton
emission in the elevated trap regime. $V_s=-2$V, $V_g=-1.5$V,
$V_t=-.5$V, $P_{ex}=750$ $\mu$W. (d) Energy of the indirect excitons
vs electrode voltage. (e) Emission spectra of indirect excitons in
the normal trap. $V_s=-2$V, $V_g=-2.5$V, $V_t=-3.5$V, $P_{ex}=18$
$\mu$W. (f) Emission spectra of indirect excitons in the normal trap
for defocused laser excitation. $V_s=0$, $V_g=0$, $V_t=-.5$V.
$P_{ex}=74$ $\mu$W defocused over 0.1 mm$^2$. (g) Ratio of line 2 to
line 1 intensities vs energy difference of the indirect excitons in
the trap $t$ and surrounding gates $g$. $P_{ex}=18$ $\mu$W. (h)
Emission spectra of indirect excitons in the elevated trap.
$V_s=-2$V, $V_g=-1.5$V, $V_t=-.5$V, $P_{ex}=18$ $\mu$W. $T=1.4$K.}
\end{center}
\end{figure}

An in-plane trap was created by a pattern of electrodes, Fig.\,1b.
An electric field $F_z$ perpendicular to the QW plane results in the
exciton energy shift $\delta E = e F_z d$, Fig.\,1d. The laterally
modulated electrode voltage $V(x,y)$ creates an in-plane relief of
the exciton energy  $E(x,y) = eF_z(x,y)d \propto V(x,y)d$
\cite{Huber98}. CQW structure was grown by MBE. $n^+$-GaAs layer
with $n_{Si}=10^{18}$ cm$^{-3}$ serves as a homogeneous bottom
electrode. The electrodes on the surface of the structure were
fabricated by depositing a semitransparent layer of Pt (8 nm) and Au
(2 nm). Two 8 nm GaAs QWs separated by a 4 nm
Al$_{0.33}$Ga$_{0.67}$As barrier were positioned 0.1 $\mu$m above
the $n^+$-GaAs layer within an undoped 1 $\mu$m thick
Al$_{0.33}$Ga$_{0.67}$As layer. Positioning the CQW closer to the
homogeneous electrode suppresses the in-plane electric field, which
otherwise can lead to the exciton dissociation \cite{Hammack06}. The
excitons were photoexcited by a 633 nm HeNe laser focused to a spot
$\sim 5 \mu$m in diameter in the area shown by the upper circle in
Fig. 1b \cite{3spots}. The emission images were taken by a CCD with
an interference filter $790 \pm 5$ nm, which covers the spectral
range of the indirect excitons. The spatial resolution was 2
microns. The spectra were measured using a spectrometer with
resolution 0.18 meV.

The electrode pattern shown in Fig.\,1b was used to create normal
traps at $|V_t| > |V_g|, |V_s|$ as well as elevated traps at $|V_t|
< |V_g|, |V_s|$ for the indirect excitons in the region of central
electrode $t$. The cloud of indirect excitons in the trap is in the
center of Fig.\,1c. In the normal trap the energy of the indirect
excitons is below the energy of the surrounding areas, $E_t < E_g,
E_s$, while in the elevated trap it is above $E_t > E_g, E_s$. For
the entire range of the studied densities ($P_{ex}= 1-1140 \mu$W
with the smallest limited by the signal strength) the emission
spectrum in the normal trap was a structureless line with FWHM $> 1$
meV, like that in Fig.\,1e. However, sharp lines were observed in
the emission spectrum in the elevated trap, Fig.\,1h. The sharp
lines emerge at the transition from the normal trap to elevated
trap, Fig.\,1g.

\begin{figure}
\begin{center}
\includegraphics[width=7cm]{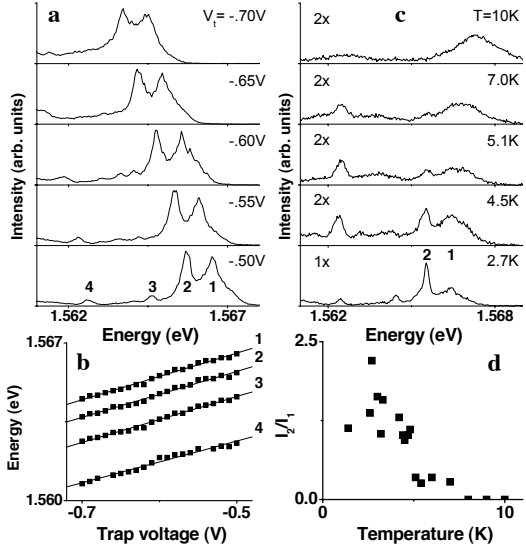}
\caption{(a) PL spectra  of excitons in the elevated trap and (b)
energies of lines 1-4 vs $V_t$. $V_s=-2$V, $V_g=-1.5$ V, $T=1.4$K,
$P_{ex}=18 \mu$W. (c) PL spectra of excitons in the elevated trap vs
bath temperature. $V_s=-2$V, $V_g=-1.5$V, $V_t=-.5$V, $P_{ex}=18
\mu$W. (d) The ratio of line 2 intensity to line 1 intensity vs bath
temperature. $V_s=-2$V, $V_g=-1.5$V, $V_t=-.5$V, $P_{ex}=6.3 \mu$W.}
\end{center}
\end{figure}

FWHM of lines 2-4 were measured as low as 0.18 meV, which is close
to the spectrometer resolution. Sharp lines in the emission of
excitons in QWs were observed in earlier studies
\cite{Zrenner94,Hess94,Gammon96}. They were attributed to the
emission of excitons localized in the local minima of the disorder
potential. The earlier reported sharp lines correspond to the
emission of direct excitons with both electrons and holes confined
in one QW. However, varying the electrode voltage $V_t$ results in
the energy shift of lines 1-4, Fig.\,2a,b. The measured energy shift
$\delta E \approx 10.4 V_t$ meV/V corresponds to $d \approx 10.7$
nm, which is close to the nominal distance between the QW centers.
Therefore, lines 1-4 correspond to the emission of the indirect
excitons.

\begin{figure*}
\begin{center}
\includegraphics[width=13.9cm]{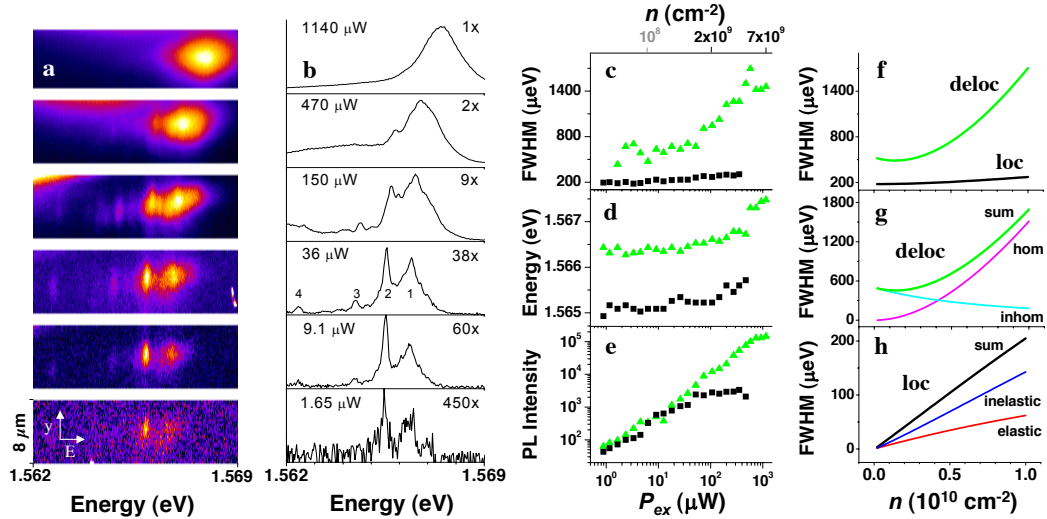}
\caption{(a) $E-y$ emission images and (b) spectra of indirect
excitons in the elevated trap vs $P_{ex}$. (c) FWHM, (d) energy, and
(e) intensity  of line 1 (green triangles) and line 2 (black
squares) vs $P_{ex}$. Density listed on top is an estimate based on
the exciton energy shift, the lowest value is an estimate based on
the PL intensity ratio. $V_s=-2$V, $V_g=-1.5$V, $V_t=-.5$V,
$T=1.4$K. (g) Calculated FWHM of emission line of delocalized
indirect excitons (green). Thin lines present the contributions from
homogeneous broadening due to exciton-exciton interaction (magenta)
and inhomogeneous broadening due to disorder (cyan). The disorder
amplitude $U^{(0)}/2=0.35$ meV corresponds to line 1. (h) Calculated
FWHM of emission line of localized indirect excitons (black). Thin
lines present the contributions from inelastic (blue) and elastic
(red) exciton scattering. $\varepsilon=1.3$ meV that corresponds to
line 2. (f) Calculated FWHM of emission lines including the
resolution 0.18 meV for comparison with the experimental data in
(c).}
\end{center}
\end{figure*}

We attribute the low energy lines 2-4 to the emission of the exciton
states localized in local minima of the disorder potential and the
highest energy line 1 to the emission of excitons delocalized over
the trap. The observation of the individual localized states in the
disorder potential is facilitated by collecting the exciton emission
from a small area, Fig.\,1c, containing not too many localized
states so that their emission can be resolved. In general, particles
in a weak disorder are delocalized while deep minima, which occur in
the disorder potential, produce strongly localized states. The
disorder potential in QW structures generally has a broad
distribution of lengths and depths
\cite{Hegarty84,Takagahara85,Zrenner94,Hess94,Gammon96,Zimmermann97}.
Localized and delocalized states can be distinguished via a
different density dependence of their emission, as described below.

Utilizing an elevated trap leads to an effective ``cooling" of the
excitons. Since the energy of excitons in the elevated trap is
higher than in the surrounding areas, excitons can escape the trap.
The escape rate of excitons with a lower energy is generally slower.
This increases the relative occupation of the lower energy states
and, therefore, can be considered as lowering the exciton
temperature. This mechanism of ``cooling'' has similarities with
evaporative cooling. The ``cooling efficiency'' can be quantified by
the relative populations of the localized and delocalized states.
One can introduce an effective temperature ${\tilde T}$ for the
population balance between these states as $\delta=n_{\rm deloc} /
n_{\rm loc} \sim \exp[{-\varepsilon/(k_{\rm B}{\tilde T}})]$, where
$\varepsilon$ is a localization energy. Then the ``cooling
efficiency'' in the elevated trap regime can be presented by the
parameter $\gamma=\delta_{\rm norm} / \delta_{\rm elev}$. A large
$\gamma$ observed in the experiments indicates an efficient cooling.
In turn, the temperatures in the elevated trap regime and in the
normal trap regime are related by $1/{\tilde T}_{\rm elev}=1/{\tilde
T}_{\rm norm} + (k_{\rm B} / \varepsilon) \ln{\gamma}$. This allows
a rough estimate for ${\tilde T}_{\rm elev}$. Assuming $\gamma
\gtrsim 10$ and ${\varepsilon/k_{\rm B}} \sim 10$K (correspond to
line 2, Fig.\,1) and ${\tilde T}_{\rm norm} \sim 3$K (corresponds to
a typical temperature of the indirect excitons in the presence of cw
excitation and without the "cooling" \cite{Butov01,Hammack06b}), one
obtains ${\tilde T}_{\rm elev} \lesssim 1.8$K.

The ``cooling efficiency'' $\gamma$ can be estimated by using the
rate equations for occupations of the localized and delocalized
states at equilibrium, which result to $\gamma = 1 + \tau /
\tau_{\rm esc}$, where $\tau^{-1} = \Lambda^{-1}\tau_{\rm
loc}^{-1}+\tau_{\rm opt}^{-1}$, $\Lambda^{-1}=(\Lambda_{\rm
l}+\Lambda_{\rm d})/\Lambda_{\rm l}$, $\Lambda_{\rm d}$
($\Lambda_{\rm l}$) is the population rate of the delocalized
(localized) states due to the laser excitation, $\tau_{\rm opt}$ and
$\tau_{\rm loc}$ are the radiative and localization times of the
delocalized excitons, and $\tau_{\rm esc}$ is their escape time from
the elevated trap ($\tau_{\rm esc} \sim S_{\rm trap}/D_{\rm x} \sim$
0.1-1$\,ns$ for the elevated trap area $S_{\rm trap} \sim 1
\mu$m$^2$, Fig.\,1c, and the diffusion coefficient $D_{\rm x} \sim
10-100$ cm$^2$/s \cite{Ivanov06}). For direct excitons with a short
lifetime $\tau_{\rm opt} \sim 0.01-0.1$ ns, the cooling is
inefficient. Indeed, in this case $\gamma < 1 + \tau_{\rm
opt}/\tau_{\rm esc} \sim 1.1$. On the contrary, the cooling can be
efficient for the indirect excitons with a long lifetime $\tau_{\rm
opt} \sim 10-10^{4}$ ns. In this case $\gamma \sim 1 + \Lambda
\tau_{\rm loc}/\tau_{\rm esc}$. It reaches $\gamma \sim 10$
estimated from the experiments when $\tau_{\rm esc} \sim 0.1 \Lambda
\tau_{\rm loc}$.

Note also that in the normal trap regime, the excitons created by
the laser excitation in the region of electrode $s$ travel to the
trap following a downhill potential energy gradient. As shown in
\cite{Hammack06b} traveling downhill may increase the exciton
temperature. However, this heating mechanism is absent in the
elevated trap regime \cite{cooling}. The lower effective exciton
temperature in the elevated trap compared to that in the normal trap
facilitates the emergence of the lower energy lines in the elevated
trap regime, Fig.\,1g,e,h. Note also that the relative intensity of
the lower energy lines 2-4 in the elevated trap regime reduces with
increasing bath temperature, Fig.\,2c,d.

We also performed an experiment at conditions in between these
regimes: We studied excitons in the normal trap using defocused
excitation, Fig.\,1f. At these conditions, a significant fraction of
the excitons is created in the trap. Therefore the heating due to
the downhill travel present in a normal trap with remote excitation
(Fig.\,1e) is reduced yet the ``cooling mechanism'' of the elevated
trap (Fig.\,1h) is absent. Sharp lines can be observed in this
regime; however, they are not as pronounced as in the elevated trap.
The rest of the paper concerns the elevated trap regime where the
sharp lines are more distinct.

The exciton energy increases with density due to the repulsive
dipole-dipole interaction of the indirect excitons, Fig.\,3d. The
exciton density $n$ can be estimated from the energy shift $\delta
E$ as $n = \epsilon \delta E /(4\pi e^{2} d)$
\cite{Yoshioka90,Zhu95,Lozovik96,Ivanov02}. It is presented at the
top in Figs. 3c-e. The energy shift of the localized states is close
to that of the delocalized states.

The emission intensity of the localized excitons saturates as shown
in Fig. 3e for line 2. The saturation indicates that only a finite a
number of excitons can occupy a local minimum of the disorder
potential. For the minimum corresponding to line 2 this number
appears to be one. Indeed, if a second exciton were added to the
minimum, the energy of the exciton state in it would increase due to
the repulsive interaction. Since the area of the minimum is small
compared to the area of the entire trap, adding one more exciton
should result to a significant increase of the exciton density in it
and, in turn, to a sharp increase of the energy of the localized
state. Since no such energy increase is observed, no more than one
exciton can occupy the local potential minimum. This is consistent
with the small estimated exciton localization length in the minimum
$l_{loc} \sim \hbar/\sqrt{(2M_x \varepsilon)} \sim 10$ nm, which
does not exceed the exciton Bohr radius. The effect is similar to
the Coulomb blockade for electrons. It is due to the dipole-dipole
repulsion of the indirect excitons.

The spectral width of the delocalized state emission is given by
$\Delta_{\rm FWHM} = [\Gamma_{\rm hom}^2 + \Gamma_{\rm
inhom}^2]^{1/2}$, where $\Gamma_{\rm hom} = \hbar / \tau_{\rm x-x} +
\hbar / \tau_{\rm x-LA} +\hbar / \tau_{\rm rec}$ is the homogeneous
broadening due to exciton -- exciton and exciton -- phonon
scattering and finite recombination lifetime and $\Gamma_{\rm inhom}
= \langle U_{\rm rand} \rangle$ is the inhomogeneous broadening due
to disorder, $\langle U_{\rm rand} \rangle$ is the average amplitude
of the CQW in-plane long-range disorder potential. For $n \gtrsim
10^{8}$\,cm$^{-2}$ relevant to the experiments, $\Gamma_{\rm hom}
\simeq \hbar/\tau_{\rm x-x}$ with $\tau_{\rm x-x}$ given by
\begin{eqnarray}
&&{1 \over \tau_{\rm x-x}} = \left( {u_0 M_{\rm x} \over 2 \pi}
\right)^{2} \, \left( { k_B T \over \hbar^{5} } \right) \ e^{-T_0/T}
\ F_0 \int_0^{\infty} du \int_0^{2 \pi} d
\phi \nonumber \\
&&\times { e^{2u} \over \big[ e^{u(1 - \cos \phi)} - F_0 \big] \big[
e^{u(1 + \cos \phi)} - F_0 \big] \big[ e^{2u} - F_0 \big] } \, ,
\label{e1}
\end{eqnarray}
where $F_0 = 1 - e^{-T_0/T}$, $T_0 = \pi \hbar^{2}n/(2M_{\rm x})$,
and $u_0$ is approximated by $u_0 \simeq 4 \pi e^{2}d/\epsilon$
\cite{Ivanov99}. $\Gamma_{\rm hom}(n)$ evaluated with
Eq.\,(\ref{e1}) is shown in Fig.\,3g.

The inhomogeneous broadening is dominant at low densities where the
interaction effects vanish. It produces a linewidth of 0.5 meV for
the studied sample, Fig.\,3c. The inhomogeneous broadening decreases
with increasing density due to screening of the long-range disorder
potential by interacting indirect excitons as
\begin{equation}
\Gamma_{\rm inhom} = { U^{(0)} \over 1 + [(2M_{\rm x})/(\pi
\hbar^{2})](e^{T_0/T} -1) u_0 } \, , \label{e2}
\end{equation}
where $U^{(0)} = 2 \langle |U_{\rm rand}({\bf r}_{\|})| \rangle$ is
the amplitude of the unscreened disorder potential, when $n
\rightarrow 0$ \cite{Ivanov02}. $\Gamma_{\rm inhom}(n)$ evaluated
with Eq.\,(\ref{e2}) is shown in Fig.\,3g.

The density increase also results in the homogeneous broadening of
the localized states, lines 2-4 (Fig. 3c), due to the scattering of
the localized exciton with delocalized excitons. It is given by
$\Gamma_{\rm hom}^{\rm (loc)} = \hbar/\tau_{\rm in} +
\hbar/\tau_{\rm el}$ with
\begin{eqnarray}
&&\! \! \! \! \! \! \! \! \! \! {\hbar \over \tau_{\rm in}} = \left(
{u_0 M_{\rm x} \over 2 \pi \hbar^{2}} \right)^{2} k_B T \ (1 + F_0)
\ F_0 \int_{2\Delta}^{\infty} du \int_0^{2 \pi} d
\phi \nonumber \\
&&\! \! \! \! \! \! \! \! \! \! \times { e^{2u} \over \big[ e^{u -
\Delta - f(u) \cos \phi} - F_0 \big] \big[ e^{u - \Delta + f(u) \cos
\phi} - F_0 \big] \big[ e^{2u} - F_0 \big] } \, , \label{e3}
\\
&&{\hbar \over \tau_{\rm el}} = \left( {u_0 M_{\rm x} \over 2 \pi
\hbar^{2}} \right)^{2} k_B T \ F_0 \int_0^{\infty} du \int_0^{2 \pi}
d \phi
\nonumber \\
&& \ \ \ \ \ \ \ \times { e^{u/\cos2 \phi} \over \big[e^{u/\cos2
\phi} - F_0 \big] \big[ e^{u\tan2 \phi} - F_0 \big] } \, ,
\label{e4}
\end{eqnarray}
where $f(u) = \sqrt{u(u - 2\Delta)}$ and $\Delta =
\varepsilon/(2k_{\rm B}T)$. The first (second) contribution is due
to the inelastic (elastic) scattering channel ``localized exciton +
delocalized exciton $\rightarrow$ delocalized (localized) exciton +
delocalized exciton''. In elastic scattering the delocalized exciton
changes its momentum by the amount which can be relaxed by the
localized exciton, $\sim 1/l_{loc}$. Both the experimental data,
Figs. 3c, and numerical evaluations with
Eqs.\,(\ref{e1}),(\ref{e3}),(\ref{e4}), Fig. 3g,h, show that the
homogeneous broadening of the localized state is smaller than that
of the delocalized state. This difference can be attributed to the
energy gap $\varepsilon$, which reduces the number of possible
states for the scattering of localized excitons. This difference in
linewidth broadening allows distinguishing the localized states from
the delocalized states.

This work is supported by ARO, DOE, NSF, WIMCS.


\begin{references}
\bibitem{3spots}
Results of experiments with three excitation spots at the three
circles in Fig. 1b were qualitatively similar.
\bibitem{cooling}
Traveling uphill may reduce the exciton temperature in the elevated
trap that is opposite to the case of the downhill traveling.
However, in the presented experiments this is inessential since the
estimated number of excitons which can overpass the potential energy
hill and reach the elevated trap is small compared to that of
excitons created in the trap due to the tails of the laser
excitation.
\end{references}
\end{document}